# Turbulent Particle Transport in Magnetized Fusion Plasma

C. Bourdelle, G.T. Hoang, X. Garbet

*Euratom-CEA Association, CEA/DSM/DRFC, CEA Cadarache
F-13108 Saint-Paul-Lez-Durance, France*

**Abstract**
The understanding of the mechanisms responsible for particle transport is of the utmost importance for magnetized fusion plasmas. A peaked density profile is attractive to improve the fusion rate, which is proportional to the square of the density, and to self-generate a large fraction of non-inductive current required for continuous operation.
Experiments in various tokamak devices (AUG, DIII-D, JET, TCV, TEXT, TFTR) have indicated the existence of an anomalous inward particle pinch. Recently, such an anomalous pinch has been unambiguously identified in Tore Supra very long discharges, in absence of toroidal electric field and of central particle source, for more than 4 minutes [1]. This anomalous particle pinch is predicted by a quasilinear theory of particle transport [2], and confirmed by non-linear turbulence simulations [3] and general considerations based on the conservation of motion invariants [4]. Experimentally, the particle pinch is found to be sensitive to the magnetic field gradient in many cases [5, 6, 7, 8], to the temperature profile [5, 9] and also to the collisionality that changes the nature of the microturbulence [10, 11, 12]. The consistency of some of the observed dependences with the theoretical predictions gives us a clearer understanding of the particle pinch in tokamaks, allowing us to predict more accurately the density profile in ITER.

**Introduction**
In tokamaks, the heat and particle transport occurs in the direction perpendicular to the nested toroidal magnetic flux surfaces. We refer to this direction as the radial direction and this is the unique dimension used in the following. Cross-field heat and particle transport in fusion plasmas is only partly caused by the collisional mechanisms described by neoclassical transport theory. The measured heat diffusivity is much higher than expected from neoclassical prediction, the difference being referred as "anomalous" transport (see for example [12]), believed to be driven by microturbulence. .
The particle transport case is somewhat different from the heat transport. Indeed, the heat source is often located in the core of the plasma making the distinction between a pinch and a diffusivity term very difficult. On the contrary, the particle source is often located in the outer edge region only. In such cases peaked density profiles are nevertheless observed, they are attributed to a particle pinch. A simple way to express both the diffusive and the pinch contributions to the particle flux, $\Gamma$, is:
$\Gamma = -D\nabla n + Vn$       (1)
Where n is the density profile, D is the diffusion coefficient and V is the pinch velocity. This formulation assumes that D and V are respectively weak functions of n and $\nabla n$. This article reviews particle transport in steady state conditions, for a review on perturbative particle transport see [Car].
A neoclassical pinch called the Ware pinch [14] has been long identified and compared with experimental observations. Indeed, for example, in [13] the relaxation of a density profile after a Deuterium pellet could be modelled with a pinch velocity of the order of the neoclassical pinch. Whereas, in [15], an additional anomalous particle pinch was needed to reproduce the perturbed experiments.

On the theory side, various mechanisms responsible for anomalous particle pinch have been identified, some due to the magnetic field gradient [16, 17, 21] other linked to the temperature gradient [2,18].

But it is only lately that the anomalous particle transport has been unambiguously proven in steady-state conditions. To prove the existence of this anomalous pinch the experiment must have no neoclassical pinch and no central particle source. These two conditions have been reached in Tore Supra and TCV plasmas [1, 6] where the density profiles remained peaked. It reinforces other experimental observations in favour of the existence of an anomalous particle pinch [7, 8,10,15, 19].

The object of this paper is to review the actual understanding of the particle pinch in tokamaks. First we will review the theoretical predictions of this pinch and present the latest unified view of these predictions. Then we will review the different experimental observations proving the existence of an anomalous particle pinch. The parametric dependences of this pinch found experimentally will be confronted with the theoretical predictions. Finally, the prediction of ITER density profile and its impact on the fusion power will be discussed.

**1. Theoretical studies of anomalous particle transport**

The particle flux ($\Gamma$) can be divided in two parts, a part generated by the neoclassical transport ($\Gamma_{neo}$) and a part generated by the microturbulence ($\Gamma_{turb}$).

$$\Gamma = \Gamma_{neo} + \Gamma_{turb} \quad (2)$$

The neoclassical part of the particle flux is well understood and easily modeled. Indeed, a pinch term, called the Ware pinch, has been identified in [14] where it is shown that, due to the conservation of canonical angular momentum, all trapped particles drift towards the magnetic axis with a radial velocity $V_{neo}$ such that:

$$\Gamma_{neo} = V_{neo} n \quad (3) \text{ with } V_{neo} = -\frac{E_\varphi}{B_\theta} \quad (4)$$

with $E_\varphi$ the toroidal electric field inducing the plasma current, $B_\theta$ the poloidal magnetic field. Other neoclassical effects other than the ware pinch, such as the neoclassical thermodiffusion, also contribute to the pinch and are modelled by neoclassical codes such as NCLASS [20].

The "anomalous" part of the flux generated by the microturbulence can be divided in diffusive and convective terms as follows: $\Gamma_{turb} = -D_{turb}\nabla n + V_{turb} n$ (5)

$D_{turb}$ is the diffusion coefficient and $V_{turb}$ is the velocity pinch due to microturbulence. Various theoretical predictions of the turbulent particle flux have been made, two main mechanisms have been identified: turbulence equi-partition and thermodiffusion.

The first mechanism is due to the existence of an inhomogeneous magnetic field in a tokamak induced both by the curvature and a parallel gradient of the field. In particular, the gradient of the magnetic field traps part of the particles in the magnetic well. The trapped particles are therefore responsible for instabilities especially sensitive to the magnetic field inhomogeneity. This mechanism is well described in [21], where it is found that the density of collisionless trapped particles is constant on hyper-surfaces defined by the conservation of the two first adiabatic invariants which leads to a density of trapped electrons proportional to $1/q$. $q$ is the safety factor and is equal to the ratio between the poloidal magnetic flux and the toroidal magnetic flux. $q$ increases from the core to the edge, therefore a density proportional to $1/q$ leads to a peaked profile. The magnetic field gradient term is in fact proportional to magnetic shear when Trapped Electron Modes (TEM) are dominant, as shown in [15, 16] where the effect of collisions and of passing ions is included. We will refer to this mechanism as the curvature pinch. The collisions can detrap some of the trapped electrons and therefore weaken the role of the curvature pinch.

The second mechanism is the thermodiffusion. It predicts a velocity pinch for sufficiently high $\eta = \nabla T/T / \nabla n/n$ [17, 22, 23, 24, 25] due to cold low velocity particles diffuse faster than hot and fast particles. It has also been observed in electron drift wave turbulence simulation in slab geometry [26].

Other analyses based on a more global approach of the drift wave equations for electrons and ions are finding that both mechanisms, curvature pinch and thermodiffusion, are inducing a particle pinch. For example, an analytical approach of the linear drift kinetic equation is presented in [4], 2D simulations in [2], computed particle trajectories in [mis], quasi-linear approach in [9] and 3D fluid model with a complementary analytical analysis in [3]. These more global approaches show that the curvature particle pinch is always directed inward, but that the "thermodiffusion" term can either be directed inward when Ion Temperature Gradient modes (ITG) are dominant and outward when TEM are dominant. In particular, the curvature pinch mechanism presented in [16] can be recovered in the limit of small electron pressure fluctuations, i.e in the case where trapped electrons behave as "test particles" in turbulence mainly due to ion modes [3]. These various effects are illustrated on figure 1 from [3] where the density profiles are the results of microturbulence simulations where the flux is fixed rather than the gradient. The particle flux is thus maintained to zero so that any density peaking is the signature of a turbulent pinch. If only the curvature terms are taken into account, a peaked profile is observed. Then when the thermodiffusion mechanism is implemented, the impact on the profile varies strongly with the ratio of electron heating versus ion heating, $P_e/P_i$. When $P_e/P_i = 1$, the density profile is similar to the curvature only case. When the ion heating dominates the electron heating the inward pinch is reinforced and the density profile is more peaked, the ITG modes are dominant. On the contrary, when the electron heating dominates, the density profile becomes hollow, the TEM dominates and the thermodiffusion term due to the TEM is stronger than the inward pinch due to the curvature.

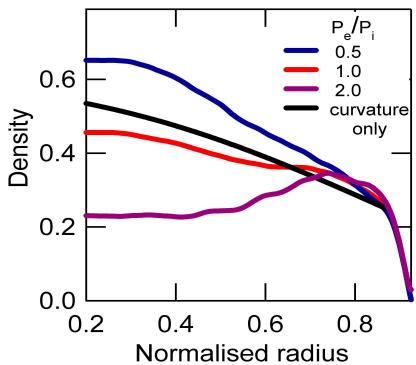

Figure 1:
Density profiles when varying the ratio of electron to ion heating, $P_e/P_i$=0.5,1 and 2, as well as the profile obtained with curvature mechanisms only.

A simple way to represent this unified theoretical view of the pinch due to drift waves in the case of no central particle source and no neoclassical pinch is to write:
$\Gamma = \Gamma_{turb} = -D_{turb}\nabla n + V_{turb} n = 0$
so: $\nabla n/n = V_{turb} / D_{turb} = -C_q \nabla q/q + C_T \nabla T_e/T_e$ (6)

$C_q$ and $C_T$ are the respective factors weighting the two contributions to the anomalous particle pinch. In this non-linear approach, the thermodiffusion can become the dominant pinch mechanism; this is not observed in a quasi-linear approach [9] where $C_q$ is found to be always of order unity whereas $C_T$ becomes similar only for specific density and temperature gradients. The fact that the non-linear simulations presented in Fig. 1 do not take the impact of collisions into account that might weaken the outward thermodiffusion term. Another open issue concerns the role of passing electrons. In the approaches presented above the passing electrons are either not taken into account [2,3], or expected to have no impact on the particle pinch [4]. But recent simulation results from [27] show that passing electrons lead to an inward pinch in cases where TEM drive an outward pinch.

## 2. Experimental evidence for anomalous particle pinch

To conclude unambiguously on the existence of an anomalous particle pinch, two important conditions have to be reached experimentally: the particle source inside a given radius must be negligible and the neoclassical pinch due to the loop voltage inducing the current must be also negligible. Under such conditions, $\nabla n/n = (V_{neo}+V_{turb}) / D$, with $V_{neo} \sim 0$, therefore the existence of a density gradient means the existence of anomalous particle pinch.

- About the particle source issue

The continuity equation is as follow:

$$\frac{\partial n}{\partial t} = -\vec{\nabla}.\vec{\Gamma} + S \qquad (7)$$

where S is the particle source. In steady state condition, $\frac{\partial n}{\partial t} = 0$, so: $\int_0^r S = \Gamma$, if the particle source inside the radius r is null, then $\Gamma = 0$, so that $\nabla n/n = (V_{neo}+V_{turb}) / D$.

In tokamaks, the source of particles has different origins. The first obvious thing to do, in order to identify the anomalous particle pinch, is to avoid any central fuelling due to the use of Neutral Beam Injection to heat the plasma and any gas puffing techniques such as pellet injection, supersonic injection, etc as well as to control the source of electrons coming from heavy ions ionization. When avoiding such conditions, the remaining particle source is due to the wall out gassing. This particle source is due to various phenomena: recycled Deuterium coming from the Carbon tiles, direct ionization of the molecular Hydrogen, molecular dissociation and creation of Franck-Condon pairs and charge-exchange of atomic Deuterium, which include molecules, atoms, charge exchange, etc. Each of these mechanisms has a penetration length that varies from less than 1 cm to up to 10 cm. Therefore larger the tokamak is, easier it is to well localize the particle source at the edge. The particle source evaluation needs Monte Carlo codes, such as [28, 29].

- About the neoclassical pinch issue

To prove unambiguously the existence of an anomalous particle pinch, an experiment has also to be performed without neoclassical pinch, so that: $\nabla n/n = V_{turb}/ D$. To reach experimentally this situation, the plasma current circulating in the torus has to be fully non-inductively driven for a time longer than the resistive diffusion time. It is only in such condition that a zero loop voltage can be reached across the whole plasma.

These experimental conditions have been reached in two tokamaks. In TCV, the full current drive situation was obtained using fast electrons from Electron Cyclotron waves (ECCD) [6] for over 4 s versus a current diffusion time of 0.4 s. In Tore Supra, the full current drive provided by the fast electrons produced by Lower Hybrid waves (LHCD) has been maintained for more than 4 minutes, more than 80 times the current diffusion time [1]. The density profile obtained in such conditions is shown on figure 2.

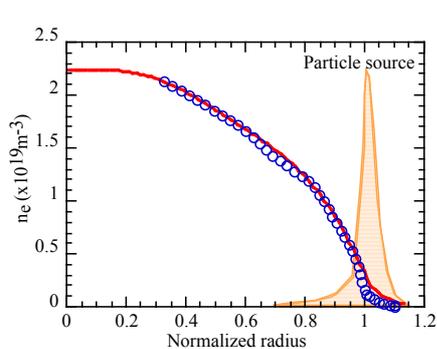

Figure 2
a. Tore Supra density profile of shot #30428 between 6 and 20 s measured by reflectometry, blue circles. The red line is the density profile reconstructed using a very simple 1D peneration model with the D and V shown in Figure 2.b. The particle source has been computed by the eirene code [28].
b. D, blue diamonds, has been taken equal to a value inferred for the ohmic phase and in general agreement with the particle confinement time of 100ms found in Tore Supra. V, red squares, matches D to reproduce the measured density profile. V is two orders of magnitude above the neoclassical pinch velocity, green triangles.

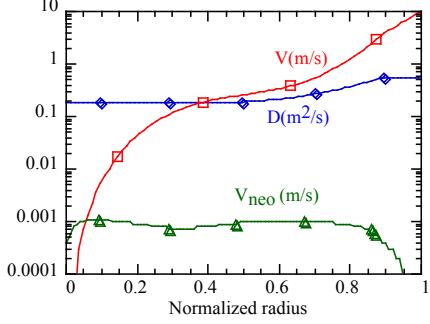

Once these experimental conditions are reached, one can deduce from the density profile peaking the ratio between $V_{turb}$ and D as follows:

$V_{turb}/D = \nabla n / n = - C_q \nabla q/q + C_T \nabla T/T$

In cases where the neoclassical Ware pinch is not zero, the analysis of experimental results may lead to contradictory conclusions. In some cases the observed pinch is found to be in agreement with $V = V_{ware}$, and D proportional to the heat diffusivity ($\chi_{eff}$) as found for JET high density H-modes [11, 30] and in ASDEX-U at high density plasmas [31, 32]. Earlier results from a perturbative particle transport study in TFTR [12] had also shown that a neoclassical pinch could explain the observed density profile relaxation. But in most other cases, the density profile cannot be explained by a neoclassical pinch alone. This is the case for JET L-modes [7, 11], DIII-D plasmas [8], TEXTOR RI modes [33] and of course for the TCV and Tore Supra results [1,6]. An anomalous particle pinch was also invoked much earlier to explain Hydrogen density modulation experiments in TEXT [15] where a velocity an order of magnitude above the neoclassical pinch velocity, scaling as 1/(nq), was needed to reproduce the observed peaked density profiles.

### 3. Parametric dependences of the anomalous particle pinch

After having demonstrated the existence of an anomalous particle pinch in section 2, the parametric dependences of this pinch are now compared to the theoretical predictions presented in section 1. In particular the predicted dependence versus $\nabla q/q$, i.e. versus the current profile shape, is tested as well as the dependence versus the temperature profiles. Finally, since the behaviors are expected to be different whereas the trapped particles are dominant or not, the dependence versus collisionality is also tested.

The particle flux is expressed as follows:

$\Gamma /n = -D(\nabla n/n + (C_q \nabla q/q - C_T \nabla T/T))+V_{neo}$

Density profile scaling as $1/q^\eta$, $\eta$ being between 0.5 and 1, has been widely tested. In TFTR, the density profiles are well fitted by the curvature pinch formula from [16]. In JET L-mode plasmas with LHCD [7], the peaking of the temperature profiles and of the current profiles are uncorrelated, therefore their respective effects on the density profile can be identified. It is found that, $ne0/<ne>$ scales as the plasma normalized internal inductance li, Fig. 3. This scaling is independent of $\nabla T_e/T_e$ and of the effective collisionality. The density profiles can be fitted using the curvature pinch formula from [16].

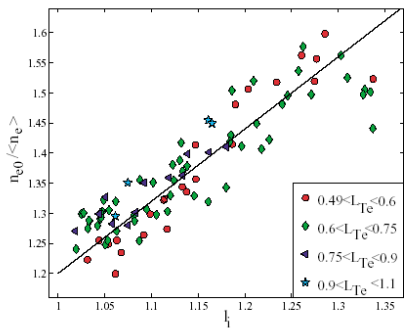

Figure 3:
Parametric dependence found for source-free Low Hybrid Current Drive L-mode JET plasma.

A linear gyrokinetic analysis done on these plasmas has been performed with GS2 [34] and shows that the unstable modes are both ITG and TEM. This could explain the lack of dependence versus $1/L_{Te}= -\nabla T_e/T_e$ evaluated at mid-radius.

In TCV [6], density profiles can be described with a curvature pinch only, with $0.4 \leq C_q \leq 1$. Most of the density profiles in Ohmic and ECH discharges, can also be reproduced with suitable combinations of the Ware pinch and the anomalous curvature pinch, assuming for instance $C_q = 0.45$ and $D/\chi_{eff} = 0.4$. In DIII-D also [8], most of the density profiles can be reproduced using an anomalous pinch term based on the adiabatic invariants approach.

There is less evidence for a correlation between the density profile and the temperature gradient, in particular because it is often difficult to decorrelate the temperature profile from the current profile. Nevertheless, a density flattening, called "pump-out", is commonly observed in response to central electron heating by ECH or LH. In some cases, the flattening can be explained by neoclassical pinches. The neoclassical thermodiffusion may be reversed if the axisymmetrical configuration is lost, leading to hollow density profiles, as seen in TCV with strong central ECH, when a saturated (1,1) island is present [36]. In other cases, the "pump-out" seems consistent with the existence of an anomalous thermodiffusion term. This is the case for the hollow density profiles observed with strong central LH in ASDEX [37]. Also, in ECH ASDEX-Upgrade plasmas (including L- and H-modes), where the nature of the unstable modes calculated with a gyrokinetic code [34] is found to be consistent with a thermodiffusion mechanism [9]. If the ITG modes are dominant, the density profile is not affected significantly by central ECH. When the density flattening is observed, $T_e$ increases at fixed $\nabla T_e/T_e$, and reinforces the TEM outward pinch.

In a non-linear microstability analysis of Alcator-C mod [38] density Internal Transport Barriers, the degradation of the confinement is associated to an outward thermodiffusion particle pinch due to the onset of TEM.

In Tore Supra, plasmas with no neoclassical Ware pinch can be studied in conditions in which: $\nabla n/n = -C_q \nabla q/q + C_T \nabla T/T$, [5]. In the core, $r/a \leq 0.30$, with a, the minor radius, $\nabla n_e/n_e$ is strongly correlated to $\nabla T_e/T_e$ with $C_T > C_q$. This inward pinch is consistent with the thermodiffusion mechanism since, using a gyrokinetic code Kinezero [39], the ITG modes are found to be dominant in this region. At mid radius, $0.35 \leq r/a \leq 0.6$, on the contrary, the density peaking is found to be weakly correlated with the temperature profile, $C_q > C_T$. At these radii, the linear gyrokinetic simulations show that the TEM are dominant.

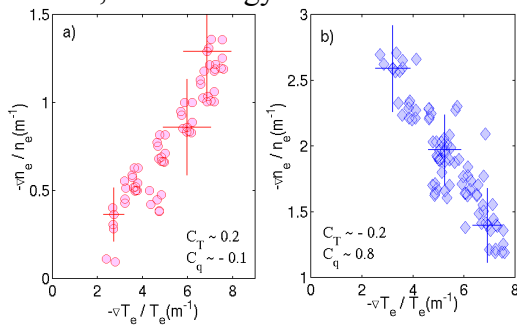

Figure 4

$\nabla n_e/n_e$ versus $\nabla T_e/T_e$ from a set of seven discharges: (a) for $r/a \leq 0.30$ with $\nabla q/q = 2 \pm 0.4$, $T_e/T_i = 2 \pm 0.4$, $\nabla T_e/\nabla T_i = 3.8 - 5$; (b) for $0.35 \leq r/a \leq 0.6$ with $\nabla q/q = 3.5 \pm 0.4$, $T_e/T_i = 1.2 \pm 0.4$, $\nabla T_e/\nabla T_i = 0.7 - 3.5$.

Collisionality is a parameter expected to have a strong impact on the density profiles, since it affects the role of trapped particles crucial for the curvature pinch [10]. The collisionality, $\nu_{eff}$, is the ratio of the collisionality detrapping the electrons over the vertical drift frequency of trapped electrons. $\nu_{eff}$ characterizes the impact of collisions on TEM. In H modes on ASDEX Upgrade [10] and on JET [30, 40], the density peaking decreases with increasing $\nu_{eff}$, Fig. 5 [Ang].

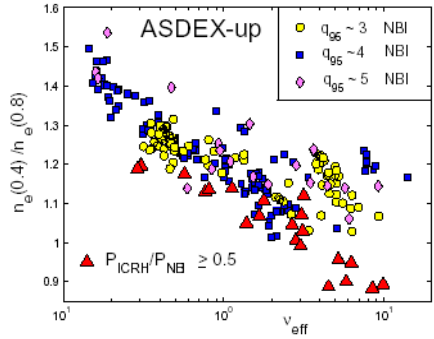

Figure 5
The ratio of the line averaged electron density at r/a = 0.4 and r/a = 0.8, $n_e(0.4)/n_e(0.8)$, is plotted versus $\nu_{eff}$.
The density peaking is strongly reduced as the collisionality increases independently of the value of q at the edge. For ITER $\nu_{eff}$ is expected to be around 0.1.

Neoclassical effects cannot explain the density peaking decrease at increasing collisionality. Indeed, a higher collisionality implies a higher resistivity, therefore a lower $E_\varphi$, hence a lower Ware pinch, see eq. (4). A detailed analysis using quasi-linear fluid drift wave modelling [10], finds the $\nu_{eff}$ dependence consistent with the observed ITG and TEM responding to the curvature pinch. Nevertheless, some puzzling paradoxes are observed. Indeed, in a curvature pinch cases, a q profile (li or $q_{95}$) sensitivity is expected. But in JET [40] and ASDEX Upgrade H modes no such dependence is observed, expect at $\nu_{eff}$ < 0.3 in JET, where the sensitivity to q profile is recovered. This unseen q profile dependence in most H modes is very difficult to understand. On the contrary, in JET L-modes [7], the density peaking strongly depends on li (Fig. 3) but is insensitive to $\nu_{eff}$. This could be explained by a q profile sensitivity mainly due to trapped ions rather than trapped electrons explaining the unseen $\nu_{eff}$ dependence.

**Discussion:**
Recently the particle transport has made progress on both experimental and theoretical sides on various tokamaks. This has led to significant progress in our understanding of the origin of the commonly observed particle pinch.

Indeed, an anomalous contribution to this pinch has been unambiguously determined in two tokamaks [1,6]. The two anomalous particle pinch contributions predicted by the theory, one scaling with the magnetic field gradient, the other scaling sensitive with the temperature gradient, are now understood as specific solutions of the general drift equations [2, 3, 4]. Still the role of passing electrons, so far mostly neglected, has to be studied carefully. Nevertheless the theoretical picture proposed in [3, 4] is consistent with several aspects of the experimental observations.

In most of the tokamak plasmas, the density profile in the gradient zone is well fitted by $1/q^n$, η ranging between 0.5 and 1 [5,7,8,16]. Near the magnetic axis, both the neoclassical pinch and the anomalous thermodiffusion pinch play a role [5,9,37]. The thermodiffusion mechanism is not clearly understood, a non-linear approach without collisions [3] predicts it to be dominant in some cases, whereas a collisional quasi-linear approach [9] predicts it to be weaker than the curvature pinch as observed experimentally. When a dependence on collisionality is observed, it is always a density peaking that decreases at higher collisionality, so opposite to the Ware pinch mechanism [40, 10]. But, at a given $\nu_{eff}$, H modes exhibit clear $\nu_{eff}$ dependence and L-modes no dependence at all [40]. These different behaviors with $\nu_{eff}$ are not yet understood.

In the light of the experimental results presented here, a flat density profile is the worst expectation for ITER. To illustrate the impact of a peaked density profile on the fusion power, ITER plasma performance has been simulated using a density profile scaling as $1/q^{0.5}$ [art] in the 0-D scaling law of the code CRONOS [41]. These simulations have been performed in a consistent manner using mixed empirical scalings and the global energy confinement scaling law ITERH-98P(y,2) [42] . An increase of the effective charge (from 1..55 to 1.7) depending on the radiated power and on electron density [43] is included. For the ITER reference scenario with 40 MW of Neutral Beam Heating, a fusion power of 530 MW (Q ~ 13) is obtained with the effect of inward curvature pinch, to be compared with 400 MW (Q =10) when using the flat density profile currently expected for ITER.

**Acknowledgements**

The author thanks G.T. Hoang, X. Garbet , Henri Weisen, Clemente Angioni, Luca Garzotti, Paola Mantica, Martin Valovic, Alexei Zabolotsky for their precious contributions to this review article.